\documentclass[pdftex,twocolumn,epjc3]{svjour3}          

\RequirePackage[T1]{fontenc}

\usepackage{ucs}
\usepackage[utf8x]{inputenc}
\usepackage{amsmath}                                 
\usepackage{array}						 

\smartqed  

\RequirePackage{graphicx}
\RequirePackage{mathptmx}      
\RequirePackage{flushend}
\RequirePackage[numbers,sort&compress]{natbib}
\RequirePackage[colorlinks,citecolor=blue,urlcolor=blue,linkcolor=blue]{hyperref}

\graphicspath{ {./images/} }

\journalname{Eur. Phys. J. C}

\begin{document}

\title{First order statistic of afterpulsing and crosstalk events in SiPMs} 


\author{S. Vinogradov 
}


\institute{P.N. Lebedev Physical Institute of the Russian Academy of Sciences, Leninskiy prospekt 53, Moscow, Russia \label{addr1} \\
\\
vinogradovsl@lebedev.ru
}

\date{Received: date / Accepted: date}

\maketitle

\begin{abstract}
This paper briefly presents an order statistic approach to the time distribution of the first detected event after a primary avalanche breakdown from a mixture of correlated and dark counting processes. The well-known order statistic method, commonly used to describe the time resolution of scintillation detectors, is applied to the arrival times of correlated events. The established model of crosstalk as a branching Poisson process is extended to afterpulsing, and correlated events are considered starting from their seeds---free (de-trapped or diffused) charge carriers capable of triggering secondary avalanche breakdowns.

The proposed approach enables the extraction of timing information for delayed crosstalk and afterpulsing events mixed with dark counts and predicts that the distribution of the first arrival time narrows as the number of seeds increases, corresponding to a higher probability of correlated events.
\end{abstract}

\section{Introduction}

Afterpulsing (AP) was first identified as a correlated noise in photomultiplier tubes (PMTs) caused by positive ion feedback in the mid-1950s \cite{breitenberger1955, Genz1968}. A decade later, AP was also observed in Geiger-mode avalanche photodiodes (APDs) and was attributed to the re-emission of charge carriers trapped in deep-level sites during a primary avalanche breakdown \cite{haitz1965, Cova1991}.

A widely used probabilistic model for AP timing describes the probability density function (PDF) of the arrival time $f(t)$ as a multi-exponential decay function \cite{Cova1991, Ohsuka1997}:
\begin{equation}
f(t)= \sum_{i} p_{i} \cdot \frac{\exp(-t/\tau_{i})}{\tau_{i}}   
\label{ExpTau}
 \end{equation}  
where the index $i$ = 1, 2... denotes each type of AP source (e.g., ions in PMTs, deep levels in APDs) with corresponding probabilities $p_{i}$ and re-emission time constants $\tau_{i}$ of presumably monoenergetic level traps.

Most studies of AP in silicon photomultipliers (SiPMs) have employed the model (\ref{ExpTau}) to extract one or two (fast and slow) time constants \cite{Du2008, Eckert2010}. 

However, experimental data on AP in InGaAs/InP Geiger-mode APDs have shown better agreement with a power function \cite{Itzler2012} or a hyperbolic sinc function \cite{Horoshko2018}. These functions result from a model of quasi-continuous deep-level distributions with constant density, contrasting with the exponential model (\ref{ExpTau}).

Recently, in addition to the de-trapping mechanism, the diffusion of minority carriers generated in an undepleted silicon bulk by secondary photons from the primary avalanche breakdown was identified as a source of delayed crosstalk (CT) and also AP due to self-triggering of the same primary fired pixel (so-called optically induced AP) \cite{Acerbi2015, Boone2017a} (see also materials from ICASiPM \cite{ICASiPM2018}).

Focusing on the specific generation and transport mechanisms responsible for the AP time profile, the aforementioned studies model AP timing under the assumption that a primary event generates a single charge carrier (seed), which then produces an AP event with some probability. This implies that the randomness of AP occurrence is governed by a Bernoulli process, while the timing randomness is governed by the seed arrival process. This approach aligns with the geometric chain process model, where a single predecessor produces either zero or one successor \cite{Vinogradov2009, Vinogradov2012NIMA}.

The similarity between delayed CT and AP processes caused by diffusion motivated the author to adopt a branching Poisson process model \cite{Vinogradov2012NIMA}, widely used for modeling the distributions of CT events \cite{Chmill2017, Vinogradov2022, Anfimov2024}. When initiated by a single primary avalanche, the branching Poisson process totally generates a random number of events of a Borel distribution, while a single branch or generation follows a Poisson distribution. It is therefore reasonable to extend the model of detected events to a random number of Poisson-distributed seeds in the first generation (de-trapped or diffused charge carriers capable of producing secondary avalanche events) with identical arrival time distributions for this generation. Each seed competes to be the first to reach the avalanche region and fire a pixel. This represents an order statistic approach, commonly used to describe the time resolution of scintillation detectors \cite{Post1950, Fishburn2010, Vinogradov2011, Vinogradov2015a}. However, unlike time resolution models that deal with observed events (fired pixels), this approach considers the seeds (potential events).

This paper presents a first-order statistic approach to the time distribution of the first detected event initiated by a random number of Poisson-distributed seeds in the first generation following a single primary event. The order statistic defines the dependence of the arrival time distribution on the number of seeds, contrasting with the geometric chain process, where the appearance of AP and its timing profile are independent. This study also unifies consideration of prompt and delayed CT and discusses similar expressions from the paper \cite{Garutti2014}.

\section{First order statistic of correlated and dark events}
The method is based on the following assumptions:
\begin{enumerate}
  \item The primary avalanche event generates a random number of seeds $N$ following a Poisson distribution with mean $\mu = E[N]$.
  \item The seed arrival times are independent and identically distributed random variables $T_{n}$ ($n$ = 1, 2 ... $N$) with the same cumulative distribution function (CDF) $F(t)= Pr(T_{n} \leq t)$ and probability density function (PDF) $f(t)=dF(t)/dt$.
  \item A correlated event is a secondary avalanche breakdown triggered by a seed in a Bernoulli process with probability $P_{trig}$.
 \item Multiple types of correlated processes with specific arrival time distributions, mean seed numbers, and triggering probabilities coexist independently. 
 \item The correlated processes are mixed with the dark counting process---an independent homogeneous Poisson point process with intensity $DCR$.
\end{enumerate}
Our objective is to derive the probability distribution of the arrival time $T_{min}$ of the first event from the mixed processes.

Consider two independent random variables $T_{C}$ and $T_{D}$. The random variable $T_{min} = \min[T_{C}, T_{D}]$ has the CDF $F_{min}(t)$:
\begin{equation}
F_{min}(t) = 1 - (1 - F_{C}(t)) \cdot (1 - F_{D}(t))  	\label{Fmin}
 \end{equation}  

If we associate $T_{D}$ with the arrival time of the first dark event from a Poisson point process with intensity $DCR$, then
\begin{align}
& F_{D}(t) = 1 - \exp(-DCR \cdot t) 		\label{Fdcr} 		\\
& F_{min}(t) = 1 - (1 - F_{C}(t)) \cdot \exp(-DCR \cdot t)	\label{FminDCR}
\end{align}  
Equation~(\ref{FminDCR}) allows extracting the probability distribution of correlated events $F_{C}$ of our interest mixed with dark counts, as demonstrated in \cite{Vinogradov2016}.
 
Now consider correlated process $C$, which generates a random number of seeds $N$ capable of producing a secondary avalanche (i.e., a correlated event). To find the marginal (unconditional) distribution of the first seed arrival time $T_{min} = \min[T_{1}, T_{2}... T_{N}]$, we sum the conditional CDFs $Pr(T_{n} \leq t | N = n)$ for a fixed $n$ over all possible values of $N$, weighted by the Poisson probability $Pr(N = n)$:
\begin{align}
F_{min}(t) & = \sum_{n=1}^{\infty} Pr(T_{n} \leq t | N = n) \cdot Pr(N = n) 			\nonumber			\\
& = \sum_{n=1}^{\infty} [1 - (1 - F_{C}(t))^{n}] \cdot \frac{\mu^{n} \cdot e^{-\mu}}{n!} \label{FminCond}
\end{align}    
The sum in Eq.~(\ref{FminCond}) can be expressed in closed form \cite{Epstein1949}:
\begin{equation}
F_{min}(t) = 1 - \exp(-\mu \cdot F_{C}(t))										\label{FminSet}
 \end{equation}  
The PDF of the distribution in Eq.~(\ref{FminSet}) is
\begin{equation}
 f_{min}(t) = \mu \cdot f_{C}(t) \cdot \exp(-\mu \cdot F_{C}(t))	\label{fminSet}
 \end{equation}  
where $\mu \cdot f_{C}(t)$ is an intensity of the inhomogeneous Poisson process of the seed arrivals.
Equations~(\ref{FminSet}) and (\ref{fminSet}) allow extracting the probability functions $F_{C}(t)$ and $f_{C}(t)$ from the dataset of $T_{n}$, as demonstrated for extracting the single-photon transit time spread function from multiphoton transit time histograms in \cite{Vinogradov2011}.

The correlated event occurs when the first seed arrives in the avalanche region and triggers a secondary avalanche with probability $P_{trig}$. This Bernoulli process is accounted for by modifying the intensity of the Poisson arrival process to $I_{C} = P_{trig} \cdot \mu \cdot f_{C}(t)$. In general, different correlated processes may have distinct triggering probabilities $P_{trig}(V_{ov})$ due to variations in the spatial localization of their sources relative to the avalanche p-n junction, as discussed in \cite{Otte2017, Gallina2019}. 

Prompt crosstalk need not be excluded from the model, even though its timing is typically unresolvable from the primary event in most practical cases. In such instances, its arrival function $f_{PC}(t)$ can be approximated by a Dirac delta function. Therefore, the total crosstalk intensity $I_{C}(t)$ is the sum of prompt $I_{PC}(t)$ and delayed $I_{DC}(t)$ components, with $P_{trig}(V_{ov}(t \rightarrow \infty)) = const$ for fully recovered pixels:
\begin{align}
 I_{C}(t) & = I_{PC}(t) + I_{DC}(t) \nonumber \\  
 & = P_{trigPC} \cdot \mu_{PC} \cdot f_{PC}(t) + P_{trigDC} \cdot  \mu_{DC} \cdot f_{DC}(t)   \label{IntensCT}
 \end{align}  

The dynamics of pixel recovery should be incorporated into the Poisson process intensity for AP events $I_{A}(t)$ via the time-dependent triggering probability $P_{trigA}(t)$:
\begin{equation}
I_{A}(t) = P_{trigA}(V_{ov}(t)) \cdot \mu_{A} \cdot f_{A}(t) 	\label{IntensAP}
 \end{equation}  
Here, $P_{trigA}(t) = P_{trigA}(V_{ov}(t))$ can be determined from experimental results or models of $P_{trig}(V_{ov})$ and $V_{ov}(t)$, as discussed elsewhere, e.g., \cite{Otte2017, Boone2017a}.

Finally, we obtain expressions for the CDF $F_{min}$ and PDF $f_{min}$ of the first event in the mixed process of dark and correlated event arrivals, defined solely by the process intensity $I_{sum}(t)$:
\begin{align}
& I_{sum}(t) = I_{A}(t) + I_{C}(t) + DCR										\label{IFin}       \\
& N_{sum}(t) = \int_{0}^{t}I_{sum}(t') dt'     							 	\label{NFin}      \\
& F_{min}(t) = 1 - \exp(-N_{sum}(t)) 						       					\label{FminFin}	\\	
& f_{min}(t) = I_{sum}(t) \cdot \exp(-N_{sum}(t))      								 \label{fminFin}
 \end{align} 

\section{Discussion}

Let us compare the geometric chain model (with event probability $p$) and the branching Poisson model (with mean number of events per branch $\mu$) in the simplest case: a single exponential distribution of arrival times $f(t)$, a single type of correlated process (delayed CT with $P_{trig} = 1$), and no dark counts ($DCR = 0$):
\begin{align}
f(t) & = \frac{e^{-t/\tau}}{\tau}						 	\label{C1} \\
F(t) & = 1 - e^{-t/\tau} 								 \label{C2} \\
f_{geom}(t) & = p \cdot f(t) = p \cdot \frac {e^{-t/\tau}}{\tau}				 \label{C3} \\
F_{geom}(t) &= p \cdot F(t) = p \cdot (1 - e^{-t/\tau})						 \label{C4} \\					
f_{poiss}(t) & = \mu \cdot f(t) \cdot e^{-\mu \cdot F(t)} = \mu \cdot \frac{e^{-t/\tau}}{\tau} \cdot e^{-\mu \cdot (1 - e^{-t/\tau})} 		  \label{C5} \\	
F_{poiss}(t) & =1 - e^{-\mu \cdot F(t)}) = 1 - e^{-\mu \cdot (1 - e^{-t/\tau})} 	 		  \label{C6}
 \end{align} 
Here, the mean number $\mu$ and the probability $p$ are related by \cite{Vinogradov2012NIMA}:
\begin{equation}
p = F_{geom}(t \rightarrow \infty) = F_{pois}(t \rightarrow \infty) = 1 - \exp(-\mu)		\label{InfP}
 \end{equation}  

For a low mean number of seeds ($\mu \ll 1$), $\mu$ and $f_{pois}(t)$ are approximately equal to $p$ and $f_{geom}(t)$, respectively:
\begin{align}
& p = 1 - \exp(-\mu) \approx \mu 						 	\label{Pmu}  \\
& f_{pois}(t) \approx \mu \cdot f_{exp}(t) \approx f_{geom} 								\label{fminLow}
 \end{align} 
In this case, the most probable non-zero number of arriving seeds from a Poisson branch is one, matching the geometric process with a single arrival seed, and the model results are similar.

Conversely, for a large mean number of seeds $\mu$, the exponential term in Eq.~(\ref{fminFin}) significantly compresses the distribution toward lower minimum values, i.e., shorter arrival times and narrower widths. This effect is well-known in multiphoton pulse arrival time measurements, where it accounts for the dependence of time walk and time resolution on the number of photons. 

\begin{figure} 
\centering
\includegraphics[width=\columnwidth]{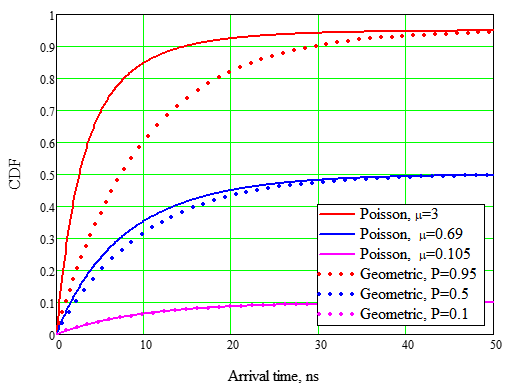}
\caption{Cumulative distribution function (CDF) of the arrival times of the first correlated event from the Poisson branching process (Eq.~(\ref{C6})) and a single correlated event from a geometric chain process (Eq.~(\ref{C4})). The Poissonian means $\mu$ are set to equalize the probabilities $p$ via Eq.~(\ref{InfP}), with an exponential time constant $\tau = 10$ ns.}
\label{1 CDF}
\end{figure}

\begin{figure} 
\centering
\includegraphics[width=\columnwidth]{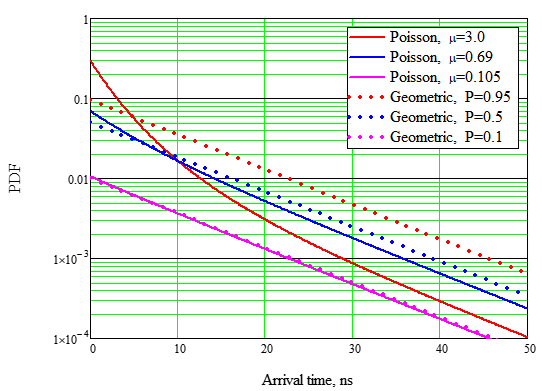}
\caption{Probability density function (PDF) of the arrival times of the first correlated event from the Poisson branching process (Eq.~(\ref{C5})) and a single correlated event from a geometric chain process (Eq.~(\ref{C3})). The Poissonian means $\mu$ are set to equalize the probabilities $p$ via Eq.~(\ref{InfP}), with an exponential time constant $\tau = 10$ ns.}
\label{2 PDF}
\end{figure}

The comparative behavior of the geometric chain model and the branching Poisson model (Eqs.~(\ref{C1})--(\ref{C6})) with equal correlated event probabilities according to Eq.~(\ref{InfP}) is shown in Figs.~\ref{1 CDF} and~\ref{2 PDF}.
 
It is worth noting that one published model of the AP time distribution \cite{Garutti2014} (see Eqs.~(3)--(5)) nearly coincides with Eqs.~(\ref{C5}) and (\ref{C6}). The authors disagreed with the exponential AP models in \cite{Du2008, Eckert2010} and proposed an alternative model applying Poisson statistics to dark counts and, without discussion, to AP events. However, the authors did not address incompatible definitions of AP probabilities in their paper. The ``total afterpulse probability'' denoted by $\epsilon_{AP}$ corresponds to the geometric model of AP events (Eqs.~(\ref{C3}) and (\ref{C4})), while the ``probability of one or more events until time $t$'' denoted by $P_{>0}(t)$ is defined for the Poisson model of dark and AP events Eq.~(\Ref{FminFin}). This duality leads to a contradiction: without dark events ($DCR=0$), the probability $P_{>0}(t)$ should describe only AP events, so $P_{>0}(t \rightarrow \infty) =  1 - \exp(-\epsilon_{AP})$. However, $P_{>0}(t \rightarrow \infty)$ should be equal to $\epsilon_{AP}$, as is the case for $F_{geom}$ and $F_{pois}$ in Eq.~(\ref{InfP}). This contradiction can be resolved by interpreting the AP probability $\epsilon_{AP}$ as the Poissonian mean $\mu$. 

Note that all expressions in this paper are derived under the assumption of point processes and should be modified to account for discrimination of events from each other and from noise. This involves attributing specific amplitude and pulse shape characteristics to events, as well as considering discrimination conditions. Under these conditions, some CT and AP pulses may be lost during an initial time interval if they are unresolved from the primary pulse or rejected due to low amplitude during pixel recovery. However, such considerations require more detailed analysis.

\section{Conclusion}
The first-order statistic approach defines the dependence of the arrival time distribution on the number of correlated seeds in a Poisson branch, contrasting with the geometric chain process, where the appearance of correlated events and their timing profiles are independently random. This dependence complicates the study of correlated processes, particularly for SiPMs with significant radiation damages.
 
\begin{acknowledgements}
The author thanks his colleagues Elena Popova and Alexander Kaminsky from CERN for discussions on experimental results for radiation damaged SiPMs.
\end{acknowledgements}
 
\bibliographystyle{spphys}
\bibliography{Bibfile_AP_EPJC}

\end{document}